\documentclass[fleqn,9pt, twocolumn]{wlscirep2}

\usepackage{graphicx}
\usepackage{dcolumn}
\usepackage{bm}
\usepackage{color}
\usepackage{siunitx}

\usepackage{amssymb,amsfonts,amsmath}

\newcommand{\DT}{\Delta T}
\newcommand{\DTc}{\Delta T_\mathrm{c}}
\newcommand{\Ttop}{T_\mathrm{top}}
\newcommand{\Tbtm}{T_\mathrm{btm}}

\title{Experimental characterization of autonomous heat engine based on minimal dynamical-system model} 

\author[a]{Shoichi Toyabe}
\author[b, *]{Yuki Izumida} 
\affil[a]{Department of Applied Physics, Graduate School of Engineering, Tohoku University, Sendai  980-8579, Japan}
\affil[b]{Department of Complexity  Science and Engineering, Graduate School of Frontier Sciences, University of Tokyo, Kashiwa  277-8561, Japan}
\affil[*]{izumida@k.u-tokyo.ac.jp}

\begin{abstract}
The autonomous heat engine is a model system of autonomous nonequilibrium systems like biological cells, exploiting nonequilibrium flow for operations.
As the Carnot engine has essentially contributed  to the equilibrium thermodynamics, autonomous heat engine is expected to play a critical role in the challenge of constructing nonequilibrium thermodynamics.
However, the high complexity of the engine involving an intricate coupling among heat, gas flow, and mechanics has prevented simple modeling.
Here, we experimentally characterized the nonequilibrium dynamics and thermodynamics of a low-temperature-differential Stirling engine, which is a model autonomous heat engine.
Our experiments demonstrated that the core engine dynamics are quantitatively described by a minimal dynamical model with only two degrees of freedom.
The model proposes a novel concept that illustrates the engine as a thermodynamic pendulum driven by a thermodynamic force.
This work will open a new approach to explore the nonequilibrium thermodynamics of autonomous systems based on a simple dynamical system.
\end{abstract}

\date{\today}

\begin{document}

\flushbottom
\maketitle

Modern physics is challenging to characterize autonomous nonequilibrium systems like biological cells.
These systems are typically complex, but there has been a long pursuit for deriving universal and simple relations governing them.
For this purpose, the extension of thermodynamics would be a promising approach because thermodynamics illustrates a universal structure of the system behind the details.
Thermodynamics is originally formulated based on infinitely slow quasistatic processes\cite{Callen1985}.
The recent challenge in constructing the finite-time thermodynamics tries to characterize the thermodynamic quantities of nonequilibrium systems by incorporating the finite-speed dynamics.
The finite-time thermodynamics was already successful in characterizing the efficiency at maximum power\cite{Curzon1975, Salamon2001,VDBroeck2005,Jimenez2007,Esposito2009,Esposito2010,Benenti2011,Izumida2014}, optimal control with the minimal energy cost\cite{Schmiedl2007, Martnez2016, Tafoya2019}, the trade-off relations between the power and efficiency\cite{Brandner2015,Shiraishi2016,Raz2016,Polettini2017,Pietzonka2018}, and stochastic heat engines\cite{SerraGarcia2016,Martinez2016,Blickle2011}.

A simple model is always the basis of scientific understanding.
Carnot's heat engine would be one of the most prominent examples in history and played an essential role in constructing thermodynamics\cite{Carnot}.
As a natural extension along with this approach, autonomous heat engines are expected to play essential roles in the development of finite-time thermodynamics.
The Carnot's engine requires control by an external agent, whereas the autonomous heat engines implement autonomous regulation like biological cells.
However, the high complexity of the engine dynamics involving a mechanical motion, heat flow, and gas flow has prevented simple modeling.

\begin{figure}[!t]
{\centering
\includegraphics[scale=0.98]{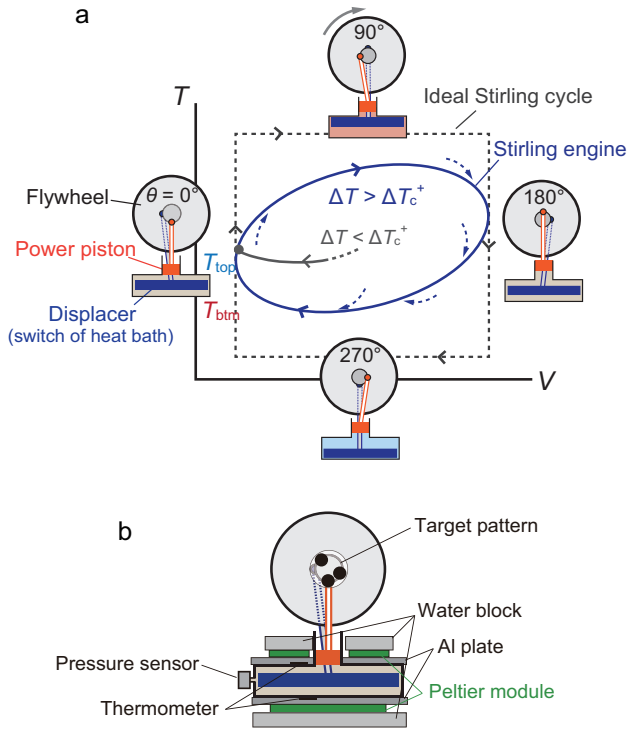}
\caption{Schematic of the experiments. {\bf a},  Thermodynamic diagram of the low-temperature-differential Stirling engine.
The heating from the bottom increases the internal pressure, pushes the power piston upward, and drives the flywheel rotation, which then pushes the displacer downward.
The displacer serves to switch the heat baths.
When the displacer moves downward, most gas in the cylinder moves to the upper side and makes contact with the top plate at a lower temperature of $\Ttop$.
The cooling of gas results in the pressure decrease and pushes the power piston downward, and the cycle restores to the initial state.
The rotation can be inverted by an opposite temperature difference.
{\bf b},  Experimental set up. 
The temperatures at the top and bottom plates are controlled by Peltier modules.
The pressure difference between the outside and inside of the cylinder is monitored by a differential pressure sensor.
The rotation of the flywheel is monitored at 100 Hz by a videoscopy of the target pattern  (three circles aligned in an isosceles triangle configuration) attached to the crank screw.
 }
\label{Fig:Intro}
}
\end{figure}

Here,  we experimentally characterize the nonequilibrium dynamics of a low-temperature-differential Stirling engine (LTD-SE), which is a model autocatalytic heat engine with a minimal structure (Fig. \ref{Fig:Intro})\cite{Senft2008}.
We especially focus on the bifurcation dynamics because the bifurcation behavior characterizes the system's universal properties behind the details.
We deduce a simple two-variable model of this engine based on the experimental results.
Such simple modeling would enable us to build a theoretical framework of autonomous heat engines towards the establishment of the finite-time thermodynamics.

The Stirling engine is an autonomous and closed heat engine \cite{Senft1993, Wolverton2008, Kongtragool2003, Koheler1995}.
Given a temperature difference, the engine cycles the volume, temperature, and pressure inside a cylinder autonomously without external timing control and rotates a flywheel unidirectionally.
Theoretically,  an ideal Stirling engine, i.e., a Stirling cycle, achieves the Carnot efficiency.
The LTD-SE \cite{Senft2008, Lu2018} consists of a power piston, displacer, flywheel, two cranks, and two rods connecting the piston and displacer to the flywheel (Fig. \ref{Fig:Intro}a).
The flywheel rotates when a sufficiently large temperature difference is given between the top and bottom plates of the cylinder.
The flywheel rotation is synchronized with the oscillation of the internal displacer and the power piston.
The displacer serves to switch the heat baths between the top and bottom plates.
Thus, the gas temperature and pressure oscillate and move the power piston up and down.
This piston motion drives the flywheel rotation.
The $\pi/2$ out of phase of the displacer and the power piston makes a cycle.
The flywheel provides inertia necessary for a smooth rotation.
When the opposite temperature difference is given, the flywheel rotates in the opposite direction with an inverted mechanism.

\section*{Experiment}

An LTD-SE (N-92 type) was bought from Kontax (UK).
We controlled the temperatures at the top and bottom plates, $\Ttop$ and $\Tbtm$, of the cylinder (Fig. \ref{Fig:Intro}b) and monitored the angular position $\theta(t)$ and angular velocity $\omega(t)$ of the flywheel and the pressure $p(t)$ inside the chamber.
See the Materials and Methods for details. 

\subsection*{Rotation}

\begin{figure}[!t]
{\centering
\includegraphics{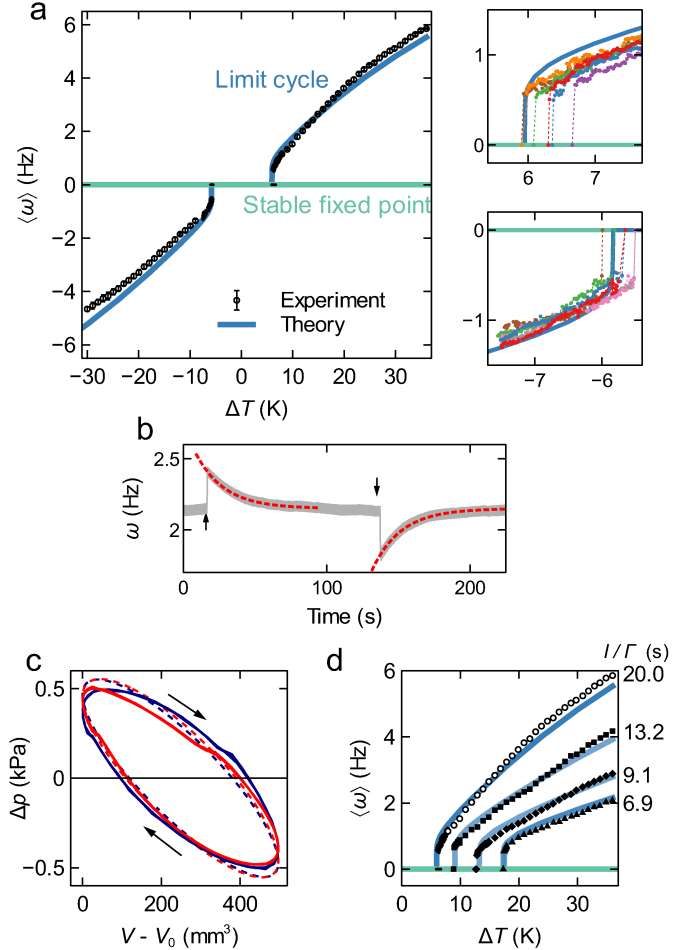}
\caption{Steady rotation. 
{\bf a}, The time-averaged angular velocity $\langle\omega\rangle$ of the limit cycle was plotted against the temperature differences $\DT$.
At $|\DT|\ge 8$K, three experimental traces were averaged (circle). The error bar corresponds to S.D. 
At $|\DT|<8$ K, twelve traces were superposed (solid lines, six for $\DT>0$ and six for $\DT<0$). 
{\bf b}, The stability of  the rotation state at $\DT=12$ K. 
The rotation state is stable against perturbations (indicated by arrows). Dashed lines are fitting curves by exponential functions, of which time constant corresponds to $I/\Gamma$.
{\bf c}, Pressure-volume curves for $\DT=12$ K (red) and -12 K (navy) obtained by experiments (solid) and theories (dashed). 
The cycling direction was clockwise independent of the sign of $\DT$.
The average of $\Delta p$ for the theoretical curves was forced to zero.
{\bf d}, Bifurcation diagrams without (open) or with (closed) additional frictional load. 
The solid curves are numerical simulations with $\Gamma$ obtained by measuring the relaxation time $I/\Gamma$ to the perturbation ({\bf b} and Fig. S2).
 }
\label{Fig:bifurcation}
}
\end{figure}

Without stimulation, the engine was settled at a stationary position $\theta \simeq -38^\circ$, where the pressure difference across the power piston and the gravity force on the power piston, displacer, crank screws, and rods are presumably balanced.
When an initial angular momentum with a sufficiently large magnitude was given, the flywheel rotated steadily with an angular velocity determined by $\DT=\Tbtm-\Ttop$ (Fig. \ref{Fig:bifurcation}a).
The rotation direction changed depending on the sign of $\DT$.
When the engine in this steady state was perturbed by hand, the angular velocity was soon recovered to the steady rate  (Fig. \ref{Fig:bifurcation}b), implying a stable limit cycle.

The pressure-volume curve exhibited a circular diagram  (Fig. \ref{Fig:bifurcation}c), demonstrating a heat engine.
The cycling direction in the PV diagram was the same independent of the sign of $\DT$, and the PV curves were nearly symmetric for the sign of $\DT$.
The area increased with $|\DT|$ (Fig. S1).

The time-averaged steady angular velocity $\langle\omega\rangle$  changed nearly linearly with $\DT$ (Fig. \ref{Fig:bifurcation}a).
$|\langle\omega\rangle|$ decreased with $|\DT|$ and vanished at a finite value of $\DT$.
The threshold value, $\DTc$, was slightly different for the sign of $\DT$; $\DTc^+=6.2\pm 0.3$ K and $\DTc^-=-5.7\pm0.2$ K  (mean $\pm$ standard deviation) , indicating the asymmetry of the dynamics for the sign of $\DT$.

The stalling at $\DTc^\pm$ was accompanied by a steep change in  $\langle\omega\rangle$, implying a homoclinic bifurcation\cite{Strogatz2014}.
The homoclinic bifurcation is a kind of a global bifurcation, and seen in, for example, a driven pendulum and a Josephson junction.
A stable limit cycle disappears with a steep but continuous transition at the homoclinic bifurcation point.
However, such the continuity is too steep to be observed in the experiments because $\omega$ is inversely proportional to $-\ln |\Delta T-\Delta T_c^\pm|$ for $|\DT|>|\DTc^\pm|$  in the vicinity of the bifurcation point\cite{Strogatz2014}.
Instead, discontinuous change in $\langle\omega\rangle$ was observed.

We also induced additional frictional load by pressing a brush for Chinese calligraphy to the flywheel.
The increase in load suppressed the angular velocity and increased $\DTc^+$ (Fig. \ref{Fig:bifurcation}c).

\subsection*{Bifurcation analysis}

\begin{figure}[!t]
{\centering
\includegraphics{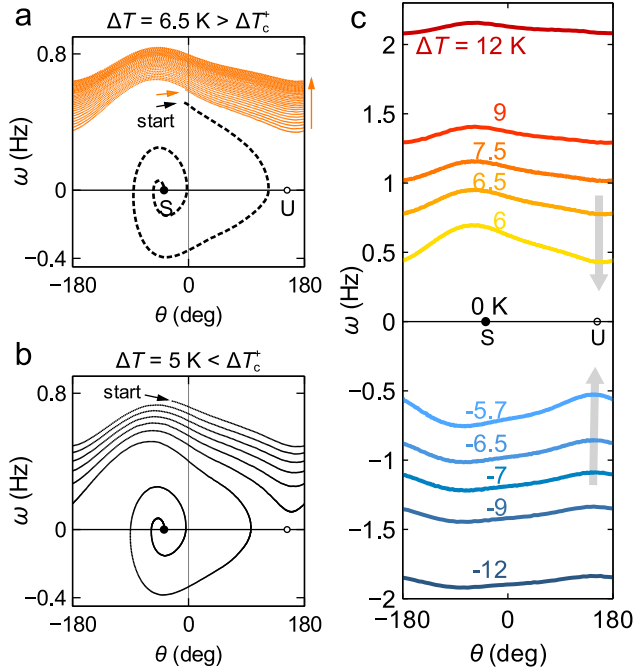}
\caption{Rotational trajectories. 
{\bf a}, Rotational trajectories at $\DT=6.5\,\mathrm{K}>\DTc^+$ initiated with small (dashed, black) or large (solid, orange) angular velocity $\omega$.
With a large initial angular velocity, $\omega$ gradually increased and, the engine settled down to a steady rotation.
With a small initial angular velocity, the engine stopped the rotation at the stable fixed point S at $(\theta, \omega) \simeq (-38^\circ, 0)$ (closed circle) after passing nearby the saddle point U at $(\theta, \omega) \simeq (153^\circ, 0)$ (open circle). 
See Fig. \ref{Fig:Intro}a for the definition of $\theta$.
{\bf b}, Rotational trajectory at $\DT=5\,\mathrm{K}<\DTc^+$.
{\bf c}, Steady rotational trajectories at different $\DT$.
 }
\label{Fig:rotation}
}
\end{figure}

We characterize the bifurcation dynamics in detail.
Figure \ref{Fig:rotation}a shows two typical trajectories started with different initial angular velocities at $\DT>\DTc^+$.
With a large initial angular velocity, we observed the convergence to the periodic trajectory determined by $\DT$.
As noted, the periodic trajectory was stable against perturbation and was identified as a stable limit cycle (Fig. \ref{Fig:bifurcation}b).

With a small initial angular velocity, the trajectory was first attracted to U at $(\theta, \omega) \simeq (153^\circ, 0)$ and then collapsed to S at $(\theta, \omega) \simeq (-38^\circ, 0)$ in a spiral-shaped manner, 
failing in converging to the stable limit cycle.
When $\DTc^-<\DT<\DTc^+$, S was the unique stable attractor (Fig. \ref{Fig:rotation}b).
These results suggest that U and S are a saddle point and a stable fixed point (spiral), respectively, and that S and the stable limit cycle coexist for $\DT>\DTc$ and $\DT<\DTc^-$.

Figure \ref{Fig:rotation}c shows the trajectories of the steady rotations, $\omega(\theta)$, at various $\DT$ above the threshold.
$\omega(\theta)$ was relatively flat at large $|\DT|$ and exhibited rugged profile at small $|\DT|$.
Specifically, as $\Delta T$ approaches $\Delta T_c$, the part of the limit cycle approaches the saddle point U, which is one of the characteristics of the homoclinic bifurcation.

All the characteristics observed above indicate the homoclinic bifurcation of the limit cycle at $\DTc$  \cite{Strogatz2014} and controvert other possibilities, including the Hopf bifurcation where local stability of the fixed point alters at the bifurcation point.
We will analyze the experimental data based on simple dynamical-system modeling below in the Theory section. 

\subsection*{Oscillatory mode}

\begin{figure}[htbp]
{\centering
\includegraphics{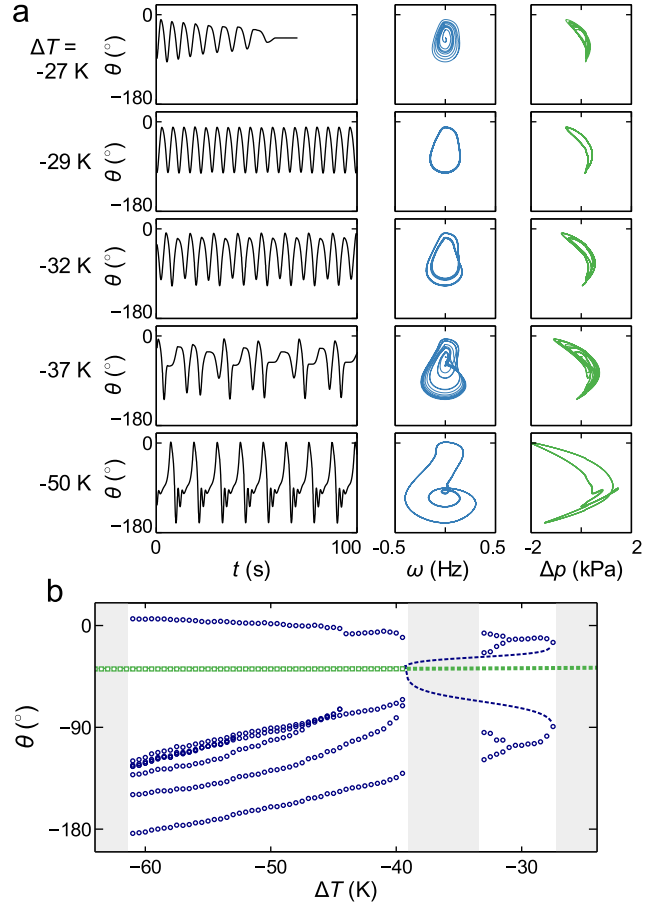}
\caption{Oscillation mode observed at $\Ttop=\SI{65}{\degreeCelsius}$.
{\bf a}, $\theta$ as the functions of time (left), $\omega$ (center), and $\Delta p$ (right). 
{\bf b}, Peak angles of the oscillation (circle), stable fixed points (closed square), unstable fixed points (open square), and expected unstable limit cycles (dashed line). 
 }
\label{Fig:oscillation}
}
\end{figure}

We also discovered an oscillation branch at $\DT\le -27$ K (Fig. \ref{Fig:oscillation}a).
Here,  for exploring a small $\DT$ region, $\Ttop$ was set to a relatively large value, 65$^\circ$C.
When we shifted the flywheel angle a little bit from S gently by hand, the flywheel started a periodic oscillation with a finite amplitude and a period of about 10 seconds, which can be considered as an oscillatory stable limit cycle.

This limit cycle showed complicated behaviors; the amplitude increases accompanied by a period-doubling bifurcation ($\SI{-27.5}{K}\ge\DT>\SI{-33.5}{K}$), seemingly aperiodic oscillation similar to chaos  ($\SI{-33.5}{K}\ge\DT>\SI{-39.5}{K}$), and again periodic oscillations accompanied by small additional oscillations  (\SI{-39.5}{K} $\ge\DT>\SI{-61.5}{K}$).
The oscillation branch disappeared at $\DT\le \SI{-61.5}{K}$, and a small perturbation got drawn into a rotation branch.
The rotation mode was observed for all $\DT<0$ with a sufficiently large initial angular velocity.
The oscillation was not observed for $\DT>0$.
The stable fixed point (spiral) at S became unstable at $\DT = -39.5$ K (Fig. S3), indicating that a subcritical Hopf bifurcation accompanied by the disappearance of an unstable limit cycle occurred. 
Although the unstable limit cycle was difficult to be identified by experiments, we may expect that the oscillatory stable limit cycle and the unstable limit cycle were created in a pair at $\DT= -27.5$ K \cite{Strogatz2014}. 
We need further studies to determine the bifurcation characteristics of the oscillation branch, including the onset and disappearance of the seemingly aperiodic oscillation.

\section*{Theoretical analysis}

For deducing the model that explains the experimental observations, we compared the above results with the theory proposed recently\cite{Izumida2018}.
The theory describes the flywheel rotation with a simple equation of motion with only two variables;
\begin{align}
\begin{split}
\dot\theta &=\omega, \\
I\dot\omega &=s\left[p(\theta,\omega)-p_0\right]r\sin\theta-\Gamma\omega. 
\end{split}\label{eq:two-variable}
\end{align}
Here, $I$ and $\Gamma$ are the moment of inertia and frictional coefficient, respectively, of the engine's rotational degree of freedom.
$r$ is the crank radius. 
$s$ is the sectional area of the power piston.
$ s[p(\theta,\omega)-p_0]\equiv s \Delta p$ corresponds to the force on the crank applied by the power piston via a rod, and $s\Delta p\cdot r\sin\theta$ is the torque on the flywheel (a piston-crank mechanism).
$p_0$ is external pressure.

The theory\cite{Izumida2018} approximates that the gas is in contact with a single heat bath at an effective temperature 
$T_0+\sin\theta\frac{\DT}2$, where $T_0 = (\Ttop+\Tbtm)/2$.
The effective temperature oscillates between $\Ttop$ and $\Tbtm$ synchronized with the displacer motion.
For the quantitative analysis of  the experimental data based on (\ref{eq:two-variable}),  we modeled the system simply as
\begin{align}
T(\theta, \omega) &=T_0+\alpha\sin(\theta-\omega\tau)\frac{\DT}2, \label{eq:T} \\
p(\theta,\omega) &=\beta \frac{nRT(\theta,\omega)}{V(\theta)}. \label{eq:P}
\end{align}
Here, the effect of the heat transfer on the gas temperature $T(\theta,\omega)$ is simply implemented by two parameters, the magnitude $\alpha$ and the time delay $\tau$ under an adiabatic assumption that the temperature equilibration is sufficiently fast compared to the flywheel dynamics.
$p(\theta,\omega)$ is calculated based on an effective equation of the state for the ideal gas.
$n$ is the amount of substance of the internal gas, and $R$ is the gas constant.
$V(\theta) =V_0+rs(1-\cos\theta)$ is the volume of the cylinder,  where  $V_0$ is the cylinder volume excluding the displacer volume.
The temperature and pressure may be nonuniform inside the cylinder, and therefore the equation of the state for the ideal gas may not hold as it is.
The coefficient $\beta$ is introduced to compensate for such an effect.

The two-variable model (\ref{eq:two-variable}) with (\ref{eq:T}) and (\ref{eq:P}) reproduced the $\DT$-dependence of $\langle\omega\rangle$ well quantitatively (Fig. \ref{Fig:bifurcation}a) including the steep change in the vicinity of $\DTc$ and the pressure-volume curve.
See Materials and Methods for the parameters used.
The model also succeeded in reproducing the bifurcation curves for the increased frictional load  (Fig. \ref{Fig:bifurcation}d).
Here, we used the same parameters except for the frictional coefficients, which were evaluated from the response curves under each condition (Fig. S2).
Note that the model  (\ref{eq:two-variable} - \ref{eq:P}) exhibits the homoclinic bifurcation as $|\DT|$ is decreased, where a stable limit cycle disappears at $\DTc^\pm$ by colliding with a saddle point at U \cite{Izumida2018}.
This is consistent with the experimental suggestions (arrows in Fig. \ref{Fig:rotation}c).
These results validate the model (\ref{eq:two-variable}).
This two-variable model is a minimal model of the autonomous heat engines in the sense that at least two variables are required to describe a limit cycle.

On the other hand, the oscillation branch (Fig. \ref{Fig:oscillation}) was not observed by this minimal model.
At $\DT\le-31.5$ K, the trajectory $\theta(\omega)$ possessed an intersection (Fig.  \ref{Fig:oscillation}a), meaning that the description by only $\theta$ and $\omega$ does no longer describe the oscillation dynamics correctly at some points.
Specifically, $p(\theta, \omega)$ was a multiple-valued function of $(\theta, \omega)$ at the intersection points, suggesting that (\ref{eq:P}) is not valid at these points.

\section*{Discussion}

An autonomous heat engine is a model system of autonomous nonequilibrium systems.
We demonstrated that the essential characteristics of the complex autonomous heat engine are reproduced by a minimal and intuitive two-variable model (\ref{eq:two-variable}) quantitatively.
The present work supports the new approach to explore the finite-time thermodynamics of autonomous heat engines based on a simple dynamical-system description.
The model contains $\DT$ explicitly through $p(\theta,\omega)$ \cite{Izumida2018}, proposing a novel concept that the LTD-SE is a thermodynamic pendulum driven by a thermodynamic force characterized by $\DT$.

Despite its simplicity, the model reproduced the essential characteristics of the engine, including the bifurcation dynamics and the thermodynamic diagram.
Whereas the model (\ref{eq:two-variable}) is derived based on the LTD-SE, this simple and intuitive formulation is expected to be applicable to a wide range of autonomous heat engines with small modifications on, for example, the cycle shape $T(\theta, \omega)$ and the piston-crank mechanism $r\sin\theta$.
The model did not reproduce the oscillation branch.
Although the oscillation is not an essential operation mode of the engine, it would be intriguing to explore what modification to the theory could successfully describe the oscillation.

The formulation of the thermodynamic efficiency of the autonomous heat engine would be of crucial importance,
which is complementary to the formulation in non-autonomous heat engines \cite{Curzon1975, Salamon2001}.
The evaluation of efficiency requires the measurement of the heat flowing through the engine and remains for future studies.

The Stirling engine is attracting growing attention in industries because it can utilize low-grade heating sources such as solar power, waste heat in the industries, and the geothermal energy, and also is environmentally friendly.
Because of its autonomous, clean, and simple machinery, the use of the Stirling engine for generating electric power for the spacecraft is being considered \cite{Wolverton2008}.
Nevertheless, the physics behind engine dynamics has been lacked.
Our experiments succeeded in characterizing the bifurcation mechanism.
Such knowledge based on physics would be effective in improving engine performance.

This work was supported by JSPS KAKENHI (18H05427 and 19K03651).

\section*{Materials and Methods}
\subsection*{Experimental setup}
An LTD-SE (N-92 type) was bought from Kontax (UK).
The temperatures of the top and bottom plates of the cylinder were controlled by Peltier modules equipped with water flowing blocks (Fig. \ref{Fig:Intro}b).
The temperatures were monitored at 2.5 Hz by Platinum resistance temperature detectors attached to the surface of the plates.
A target pattern (three circles aligned in an isosceles triangle configuration) was attached to the crank screw connected to the displacer for monitoring the angular position of the flywheel (Fig. \ref{Fig:Intro}b).
The image of the target pattern was recorded by a high-speed camera (Basler, Germany) at 100 Hz and analyzed in real time to obtain the angular position and the angular velocity of the flywheel.
A pressure sensor (Copal electronics, Japan) was fixed at the side of the cylinder to monitor the inner pressure.
We monitored the angular position $\theta(t)$, angular velocity $\omega(t)$ of the flywheel, and the pressure $p(t)$ inside the chamber under controlled $\DT(t)$.
All the experiments were controlled by a computer equipped with a program developed on LabVIEW (National Instruments).

\subsection*{Bifurcation dynamics}
For evaluating $\DT$ dependence of the angular velocity (Fig. \ref{Fig:bifurcation}a), we manually provided an initial angular momentum at $\DT$ = 36 K or -30 K with keeping $\Ttop=\SI{24}{\degreeCelsius}$ and waited for about one hour for the sufficient relaxation of the temperatures and flywheel rotation.
Then, with keeping $\Ttop=\SI{24}{\degreeCelsius}$, we varied $\DT$ from 36 K to 0 K or from -30 K to 0 K in a stepwise manner at a rate of $\pm$1 K every 180 s for $|\DT|> 8$ K and $\pm$0.02 K every 60 s or 120 s otherwise. 

\subsection*{Parameters for theoretical curves}
We used the following parameters for the theoretical curves in Figs. \ref{Fig:bifurcation}a, c, and d.
$V_0=\SI{44900}{mm^3}$, $s=\SI{71}{mm^2}$, $r=\SI{3.5}{mm}$, $I=5.7\times 10^{-5}\,\si{kg.m^2}$, and $p_0=\SI{101.3}{kPa}$. $n=\SI{0.00185}{mol}$, $R=\SI{8.314}{J /K mol}$.
We determined $\alpha$,  $\beta$, and $\tau$ as 0.17, 0.94, and 15 ms, respectively, by fitting.
The friction coefficient $\Gamma$ was measured by evaluating the relaxation time after a perturbation (Fig. \ref{Fig:bifurcation}b and S2). The relaxation time constant is approximately given by $I/\Gamma$.

\bibliographystyle{naturemag}
{\footnotesize
\bibliography{manuscript}
}

\clearpage

\renewcommand{\thefigure}{S\arabic{figure}}
\setcounter{figure}{1}

\section*{Supplementary figures}

\begin{figure}[!h]
\begin{center}
\includegraphics{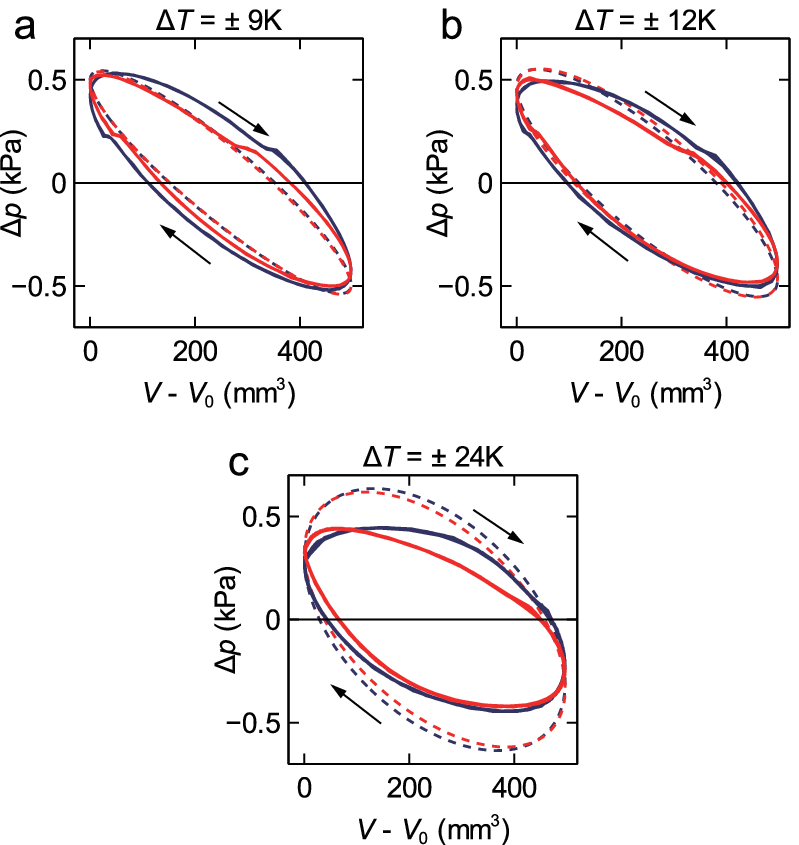}
\end{center}
\caption{Pressure-volume curves at $\Delta T= \pm 9 K$ ({\bf a}), $\pm 12$ K ({\bf b}, same as Fig. 2c), and $\pm 24$ K ({\bf c}).
 Red and navy curves correspond to $\Delta T>0$ and $\Delta T<0$, respectively.
 Solid and dashed curves correspond to experimental data and simulation data, respectively.
 }
\label{Fig:PV}
\end{figure}

\begin{figure}[!h]
\begin{center}
\includegraphics{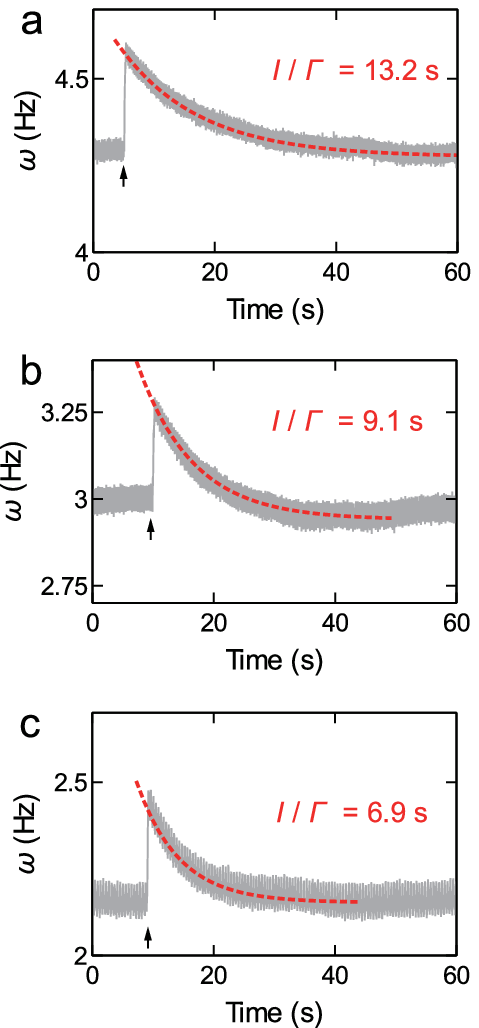}
\end{center}
\caption{The relaxation after a perturbation at a steady rotation at $\Delta T = 36$ K, corresponding to the three curves under loaded conditions in Fig. 2d.
Red dashed lines are exponential fitting.
 }
\label{Fig:relaxation}
\end{figure}

\begin{figure}[!h]
\begin{center}
\includegraphics{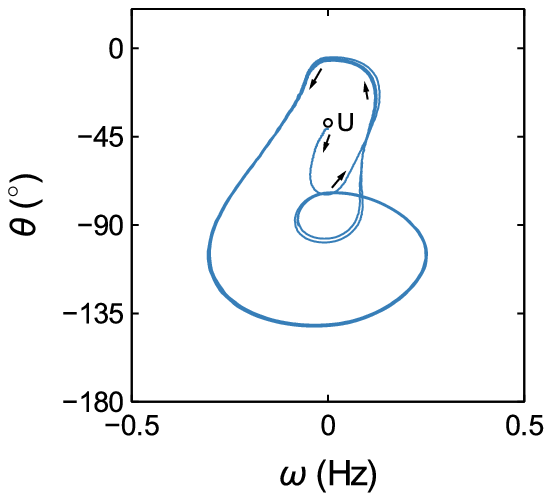}
\end{center}
\caption{When started in the vicinity of the unstable fixed point U, the trajectory converges to the oscillatory stable limit cycle in a spiral manner.
$T_\mathrm{top}=\SI{65}{\degreeCelsius}$ and $\Delta T=\SI{-42}{K}$.
 }
\label{Fig:relaxation}
\end{figure}

\end{document}